\documentclass[12pt]{article}
\usepackage{amsmath}               
\usepackage{amssymb}
\usepackage{sint}

\def\draftversion{N}                       
\def\note[#1]#2{\message{(#1)}\if\draftversion Y{\noindent\em #2}\fi}
\if \draftversion Y
\fi
\def\spose#1{\hbox to 0pt{#1\hss}}
\def\ltapprox{\mathrel{\spose{\lower 3pt\hbox{$\mathchar"218$}}%
 \raise 2.0pt\hbox{$\mathchar"13C$}}}
\def\gtapprox{\mathrel{\spose{\lower 3pt\hbox{$\mathchar"218$}}%
 \raise 2.0pt\hbox{$\mathchar"13E$}}}
\def\inapprox{\mathrel{\spose{\lower 3pt\hbox{$\mathchar"218$}}%
 \raise 2.0pt\hbox{$\mathchar"232$}}}

\def\lvec#1{\setbox0=\hbox{$#1$}
   \setbox1=\hbox{$\scriptstyle\leftarrow$}
    #1\kern-\wd0\smash{
   \raise\ht0\hbox{$\raise1pt\hbox{$\scriptstyle\leftarrow$}$}}
   \kern-\wd1\kern\wd0}

\newcommand{\half}{\frac{1}{2}}

\newcommand{\psib}{{\bar \psi}}
\newcommand{\chib}{{\bar \chi}}
\newcommand{\rhob}{{\bar \rho}}
\newcommand{\zetab}{{\bar \zeta}}
\newcommand{\rhop}{{\rho^\prime}}
\newcommand{\zetap}{{\zeta^\prime}}
\newcommand{\rhobp}{{\bar \rho^\prime}}
\newcommand{\zetabp}{{\bar \zeta^\prime}}
\newcommand{\Gx}{{\Gamma_\xi}}
\newcommand{\Gdx}{{\Gamma^\dagger_\xi}}
\newcommand{\GS}{{\Gamma_S}}
\newcommand{\GF}{{\Gamma_F}}
\newcommand{\GFT}{{\Gamma_F^T}}
\newcommand{\tOne}{{\tilde \mathbf{1}}}
\newcommand{\tG}{{\tilde \Gamma}}
\newcommand{\tGt}{{\tilde \Gamma_4}}
\newcommand{\tGf}{{\tilde \Gamma^5}}
\newcommand{\tGft}{{\tilde \Gamma^5_4}}
\newcommand{\vx}{{\vec x}}
\newcommand{\vy}{{\vec y}}
\newcommand{\Tp}{{T^\prime}}
\newcommand{\vxi}{{\vec \xi}}
\newcommand{\va}{{\vec \alpha}}
\newcommand{\vb}{{\vec \beta}}
\newcommand{\veta}{{\vec \eta}}
\newcommand{\cD}{{\mathcal D}}
\newcommand{\cO}{{\mathcal O}}
\newcommand{\hk}{{\hat k}}
\newcommand{\teta}{{\tilde \eta}}

\begin{document}

\vskip 1cm

\title{The Schr\"odinger functional running coupling with staggered fermions}

\author{Urs~M.~Heller \\[1ex]
  Supercomputer Computations Research Institute, \\
  Florida State University, Tallahassee, Florida 32306-4052, U.S.A.}
\date{May 8, 1997}

\maketitle

\vskip -8cm
\rightline{FSU-SCRI-97-41}
\vskip 7.5cm

\begin{abstract}
We discuss the Schr\"odinger functional in lattice QCD with staggered
fermions including its order $\cO(a)$ boundary counterterms. We relate it,
in the classical continuum limit, to the Schr\"odinger functional as
obtained in the same limit with Wilson fermions. We compute the strong
coupling constant defined via the Schr\"odinger functional with staggered
fermions at one loop and show that it agrees with the continuum running
coupling constant in the Schr\"odinger functional formalism.
\end{abstract}

\section{Introduction}
\label{introduction}

{}From the pioneering work of the ALPHA collaboration \cite{SF_1-SF_Z} is
has become clear that the Schr\"odinger functional in lattice QCD is a
useful setup and tool for nonperturbative computations in QCD. Through the
boundary conditions in the Schr\"odinger functional one can introduce, in a
gauge invariant way, non-vanishing background fields and use those to probe
the model. The response to a constant chromoelectric background field, for
example, allows the nonperturbative computation of a well--defined
renormalized running strong coupling constant with the scale given by the
spatial size of the system \cite{SF_1-SF_3,SF_SW}. The Schr\"odinger
functional has also proved useful for the nonperturbative computation of
the ``clover'' coefficient for $\cO(a)$ improved Wilson fermions
\cite{SF_cSW} and the gauge independent computation of current
renormalization constants \cite{SF_Z}.

In this paper we shall discuss the formulation of the Schr\"odinger
functional with staggered fermions. The basic steps have already been done
by Miyazaki and Kikukawa \cite{SF_KS}. We shall extend their work in
several directions: discuss the form of $\cO(a)$ boundary counterterms,
compute the one--loop fermionic contribution to the pure gauge $\cO(a)$
boundary counterterm ${\rm tr}(F_{k4} F_{k4})$, and most importantly,
discuss the relation between the lattice Schr\"odinger functional with
staggered fermions and the one with Wilson fermions \cite{SF_Sint} in the
continuum limit. In particular we will compute the contribution from the
staggered fermions to the running coupling constant defined through the
Schr\"odinger functional at one loop and show that it agrees with the
computations using other regularization schemes. As a by-product of this
computation we obtain, we believe for the first time, an explicit
calculation, at one loop, of lattice artefacts from staggered fermions in a
physical observable.

In the next section we will briefly review the formulation of the
Schr\"odinger functional with staggered fermions. We shall consider
massless fermions, and discuss inclusion of a mass term later in section
\ref{Schroed_m}. In section \ref{Dirac} we will discuss the Schr\"odinger
functional in the continuum limit in terms of four flavors of Dirac
fermions. In section \ref{coupling_m0} we will compute the one--loop
contribution from the staggered fermions to the Schr\"odinger functional
running coupling for massless fermions. In section \ref{Schroed_m} we
discuss inclusion of a mass term and compute the one--loop contribution to
the coupling from massive fermions. Section \ref{conclusions} contains some
concluding remarks. The symmetry properties of staggered fermions are
briefly reviewed in appendix A, and a ``two--time--slice'' transfer matrix
for staggered fermions is sketched in appendix B.

\section{The Schr\"odinger functional with staggered fermions}
\label{stag_Schroed}

The Schr\"odinger functional describes the evolution of a state at
(Euclidean) time $t=0$ to another state at time $t=T$. Using the transfer
matrix it can be written as a path integral with fixed boundary conditions
at time $t=0$ and $T$. For fermions, described classically by a first order
differential equation, actually only half the degrees of freedom can be
specified at each boundary \cite{SF_Sint,SF_Sym}. The transfer matrix for
staggered fermions has been worked out in \cite{KS_ferm}. It is non
positive, but its square is positive. This leads to a doubling of degrees of
freedom and it is better to think of this square as a transition amplitude
from two time-slices to two time-slices (see appendix B for a
``two--time--slice'' transfer matrix that corresponds explicitly to the
square of the transfer matrix derived in \cite{KS_ferm}). As shown in
\cite{SF_KS} this leads to the fact that for staggered fermions all degrees
of freedom can be fixed at both boundaries. The Schr\"odinger functional
can thus be represented as the path integral
\begin{equation}
{\mathcal Z}[W,\zeta,\zetab; W^\prime,\zetap,\zetabp] = \int [DU] \int
 \prod_{\vx}  \prod_{x_4=1}^{T-1} \left[ d\chib(\vx,x_4)
 d\chi(\vx,x_4) \right]  \exp\{-S_G - S_{SF}\} .
\end{equation}
Here $W$ and $W^\prime$ represent the boundary values of the gauge fields,
$S_G$ is the pure gauge action and $[DU]$ the Haar measure over gauge
fields, both appropriate for the Schr\"odinger functional formulation as in
\cite{SF_1}. The fermionic part of the action is given by \cite{SF_KS}
\begin{equation}
S_{SF} = \sum_{\vx} \sum_{x_4=1}^{T-1} \sum_\mu \half \eta_\mu(x)
 \chib(x) \left[ U_\mu(x) \chi(x+\mu) - U^\dagger_\mu(x-\mu)
 \chi(x-\mu) \right] + S_B^{(0)} + S_B^{(T)}
\label{eq:S_SF_chi}
\end{equation}
with $\eta_\mu(x)=(-1)^{\sum_{\nu < \mu} x_\nu}$ the usual staggered phase
factors. At the boundaries the fields take on their boundary values
\begin{equation}
\chi(\vx,0) = \zeta(\vx) , \quad
\chib(\vx,0) = \zetab(\vx) , \quad
\chi(\vx,T) = \zetap(\vx) , \quad
\chib(\vx,T) = \zetabp(\vx) ,
\label{eq:BC_chi}
\end{equation}
with $\zeta$, $\zetab$, $\zetap$ and $\zetabp$ independent complex
Grassmann fields. The additional boundary terms of the action are
\begin{eqnarray}
S_B^{(0)} &=& \sum_{\vx} \sum_{k=1}^3 \half \eta_k(\vx,0)
 \zetab(\vx) \left( W_k(\vx) \zeta(\vx+\hk) - W^\dagger_k(\vx-\hk)
 \zeta(\vx-\hk) \right) \nonumber \\
&+& \sum_{\vx} \half \eta_4(\vx,0) \zetab(\vx) \chi(\vx,1) .
\end{eqnarray}
and
\begin{eqnarray}
S_B^{(T)} &=& \sum_{\vx} \sum_{k=1}^3 \half \eta_k(\vx,0)
 \zetabp(\vx) \left( W^\prime_k(\vx) \zetap(\vx+\hk) -
 W^{\prime \dagger}_k(\vx-\hk) \zetap(\vx-\hk) \right) \nonumber \\
&-& \sum_{\vx} \half \eta_4(\vx,T) \zetabp(\vx) \chi(\vx,T-1) .
\end{eqnarray}
where $W_k$ and $W^\prime_k$ are the gauge fields at the boundaries.

Note that for staggered fermions the total number of time-slices has to be
even. Therefore, in the labeling adopted in this paper for the lattice
sites, which agrees with that of \cite{SF_1,SF_SW}, the time extent $T$ has
to be {\it odd}. In the spatial direction we take the lattice to be of size
$L$ (even!) and impose the generalized periodic boundary conditions
\cite{SF_SW}
\begin{equation}
\chi(x+L\hk) = {\rm e}^{i \theta_k} \chi(x) , \quad
\chib(x+L\hk) = \chib(x) {\rm e}^{-i \theta_k} .
\end{equation}
These boundary conditions are easily implemented by transforming to
periodic fermion fields in a constant abelian background field $u_4 = 1$,
$u_k = {\rm e}^{i \theta_k/L}$.

\section{The Schr\"odinger functional in terms of four-component spinors}
\label{Dirac}

We construct four-component spinors from the one-component Grassmann fields
$\chi$ and $\chib$ in the standard way, following  Kluberg-Stern {\it et
al.} \cite{Kluberg}. We discuss here only free staggered fermions.
Including the coupling to gauge fields is quite straightforward, though
notationally somewhat cumbersome \cite{Kluberg,Jolicoeur}. Since we will be
interested in the classical continuum limit, we shall display the lattice
spacing explicitly (except in the labelling of the lattice sites). We
divide the lattice into $2^4$ hypercubes in which the four-component
spinors reside.\footnote{This division of the lattice into $2^4$ hypercubes
requires that in each direction we have an even number of sites. With our
labeling convention in the time direction $T$ is odd and $0 \leq x_4 \leq
T$.} Thus we set $x = 2y + \xi$ with $\xi_\mu = 0, 1$ and define
\begin{equation}
\chi(2y+\xi) = \chi_\xi(y) ~, \quad \chib(2y+\xi) = \chib_\xi(y) .
\end{equation}
We also introduce
\begin{equation}
\Gx = \gamma_1^{\xi_1} \gamma_2^{\xi_2} \gamma_3^{\xi_3}\gamma_4^{\xi_4} .
\end{equation}
We use hermitian Euclidean Dirac matrices. The $\Gx$ matrices satisfy
the orthogonality and completeness relations
\begin{equation}
{\rm tr} \left( \Gdx \Gamma_{\xi^\prime} \right) = 4 \delta_{\xi \xi^\prime} ,
 \qquad  \sum_\xi \left( \Gdx \right)^{a \alpha} \Gx^{\beta b} =
 4 \delta^{a b} \delta^{\alpha \beta} .
\end{equation}
The four flavors of four-component Dirac spinors are then constructed as
\begin{equation}
\psi^{\alpha a}(y) = \frac{1}{8} \sum_\xi \Gx^{\alpha a}
 \chi_\xi(y) , \quad
\psib^{a \alpha}(y) = \frac{1}{8} \sum_\xi \chib_\xi(y)
 \left( \Gdx \right)^{a \alpha} .
\label{eq:psi_def}
\end{equation}
where greek superscripts denote spin indices and roman superscripts flavor
indices. These relations can be inverted:
\begin{equation}
\chi_\xi(y) = 2 \sum_{\alpha a}  \left( \Gdx  \right)^{a \alpha}
 \psi^{\alpha a}(y) , \quad
\chib_\xi(y) = 2 \sum_{\alpha a} \bar \psi^{a \alpha}(y)
 \Gx^{\alpha a} .
\label{eq:psi_to_chi}
\end{equation}

It is useful to introduce some more notation. Let $\Lambda = \GS \otimes
\GFT$ for some gamma matrix $\GS$ acting in Dirac space and some gamma
matrix $\GF$ acting in flavor space. Then
\begin{eqnarray}
\Lambda_1 \cdot \Lambda_2 &=& (\Gamma_{S_1} \Gamma_{S_2}) \otimes
(\Gamma_{F_2} \Gamma_{F_1})^T \nonumber \\
(\Lambda \cdot \Gx)^{\alpha a} &=& (\GS \Gx \GF)^{\alpha a} \\
(\Gdx \cdot \Lambda)^{a \alpha} &=& (\GF \Gdx \GS)^{a \alpha} \nonumber .
\end{eqnarray}
In the second and third equations above, just as in the definition of the
$\psi$-fields, (\ref{eq:psi_def}), $\Gx$ is a ``mixed'' matrix with the
first index a spinor index, and the second one a flavor index. We also
define
\begin{equation}
\tOne = {\bf 1} \otimes {\bf 1} , \quad
\tG_\mu = \gamma_\mu \otimes {\bf 1} , \quad
\tGf_\mu = \gamma_5 \otimes (\gamma_5 \gamma_\mu)^T
\end{equation}
and the projectors
\begin{eqnarray}
P_0^{(\mu)} &=& \half \left[ {\bf 1} \otimes {\bf 1} + (\gamma_\mu \gamma_5)
 \otimes (\gamma_5 \gamma_\mu)^T \right] = \half \tG_\mu \left( \tG_\mu +
 \tGf_\mu \right) , \nonumber \\
P_1^{(\mu)} &=& \half \left[ {\bf 1} \otimes {\bf 1} - (\gamma_\mu \gamma_5)
 \otimes (\gamma_5 \gamma_\mu)^T \right] = \half \tG_\mu \left( \tG_\mu -
 \tGf_\mu \right) .
\label{eq:proj_mu}
\end{eqnarray}
These projectors are useful since
\begin{equation}
P_0^{(\mu)} \cdot \Gx = \delta_{\xi_\mu,0} \Gx , \quad
\Gdx \cdot P_0^{(\mu)} = \delta_{\xi_\mu,0} \Gdx , \quad
P_1^{(\mu)} \cdot \Gx = \delta_{\xi_\mu,1} \Gx , \quad
\Gdx \cdot P_1^{(\mu)} = \delta_{\xi_\mu,1} \Gdx .
\end{equation}
Thus they project onto the one-component fields with $\xi_\mu = 0$ or 1.

Inserting the relations (\ref{eq:psi_to_chi}) into the free staggered
action, it can be written as
\begin{equation}
S_F = (2a)^4 \sum_{y,\mu} \psib(y) \left[ \tG_\mu D_\mu \psi(y) +
 \tGf_\mu a \Delta_\mu \psi(y) \right]
\label{eq:S_F_psi}
\end{equation}
where
\begin{eqnarray}
D_\mu f(y) &=& \frac{1}{4a} \left[ f(y+\mu) - f(y-\mu) \right] \nonumber \\
\Delta_\mu f(y) &=& \frac{1}{4a^2} \left[ f(y+\mu) - 2f(y) +
 f(y-\mu) \right] .
\end{eqnarray}
The expression for the action can be compactified further by defining
$\cD_\mu$ as
\begin{equation}
\cD_\mu \psi(y) = ({\bf 1} \otimes {\bf 1}) D_\mu \psi(y) +
 (\gamma_\mu \gamma_5) \otimes (\gamma_5 \gamma_\mu)^T a \Delta_\mu \psi(y) .
\label{cD_def}
\end{equation}

To deal with the boundaries in the Schr\"odinger functional we note that
the projectors $P_{0,1} \equiv P^{(4)}_{0,1}$ project onto the boundary
fields
\begin{equation}
P_0 \psi(\vy,0) = \rho(\vy) , \quad
\psib(\vy,0) P_0 = \rhob(\vy) , \quad
P_1 \psi(\vy,\Tp) = \rhop(\vy) , \quad
\psib(\vy,\Tp) P_1 = \rhobp(\vy) ,
\label{eq:BC_hyp}
\end{equation}
where we set $\Tp = (T-1)/2$ for the upper boundary hypercubes (recall that
$T$ has to be odd). The boundary four-component spinors $\rho$ and $\rhop$
are related to the boundary one-component spinors $\zeta$ and $\zetap$ as
in eq.~(\ref{eq:psi_def}),
\begin{equation}
\rho^{\alpha a}(\vy) = \frac{1}{8} \sum_\vxi 
 \Gamma_{(\vxi,0)}^{\alpha a} \zeta_\vxi(\vy) , \quad
\rhop^{a \alpha }(\vy) = \frac{1}{8} \sum_\vxi 
 \Gamma_{(\vxi,1)}^{a \alpha } \zetap_\vxi(\vy)
\label{eq:eta_def}
\end{equation}
and analogously for $\rhob$ and $\rhobp$. The action appropriate for the
Schr\"odinger functional can now be written as \cite{SF_KS}
\begin{equation}
S_{SF} = (2a)^4 \sum_{\vy} \sum_{y_4=1}^{\Tp-1} \sum_\mu \psib(y) \tG_\mu
 \cD_\mu \psi(y)
 + S_B^{(0)} + S_B^{(T)}
\label{eq:S_SF_psi}
\end{equation}
with the boundary contributions
\begin{eqnarray}
S_B^{(0)} &=& (2a)^4 \sum_{\vy} \sum_{k=1}^3 \psib(\vy,0) \tG_k
 \cD_k \psi(\vy,0)
 + (2a)^3 \sum_{\vy} \psib(\vy,0) P_1 \tGt \psi(\vy,1) \nonumber \\
&-& (2a)^3 \sum_{\vy} \psib(\vy,0) \tGft \psi(\vy,0)
\end{eqnarray}
and
\begin{eqnarray}
S_B^{(T)} &=& (2a)^4 \sum_{\vy} \sum_{k=1}^3 \psib(\vy,\Tp) \tG_k
 \cD_k \psi(\vy,\Tp)
 - (2a)^3 \sum_{\vy} \psib(\vy,\Tp) P_0 \tGt \psi(\vy,\Tp-1)  \nonumber \\
&-& (2a)^3 \sum_{\vy} \psib(\vy,\Tp) \tGft \psi(\vy,\Tp) .
\end{eqnarray}
Using $\tGft = P_1 \tGt P_0 - P_0 \tGt P_1$ we see that the last term in
both $S_B^{(0)}$ and in $S_B^{(T)}$ involves a boundary field, and hence
vanishes for homogeneous boundary conditions.

Miyazaki and Kikukawa already addressed the question as to whether there
are additional boundary counterterms contributing in the continuum limit.
They would have to be operators of dimension three, {\it i.e.} of the form
\begin{equation}
\Delta S_B = (2a)^3 \sum_{\vy} \sum_i \left\{ c_i^{(0)} \psib(\vy,0)
 \Lambda_i \psi(\vy,0) + c_i^{(T)} \psib(\vy,\Tp) \Lambda_i
 \psi(\vy,\Tp) \right\} .
\end{equation}
Taking into account the discrete spatial rotational symmetry, parity ---
the projectors $P_{0,1}$ appearing in the boundary conditions
(\ref{eq:BC_hyp}) are invariant under parity --- and the chiral U(1)
symmetry of massless staggered fermions, Miyazaki and Kikukawa concluded
that \cite{SF_KS}
\begin{equation}
\Lambda_i = \tGt , ~~ \tGft , ~~ (\gamma_4 \otimes (\gamma_4 \bar \gamma)^T)
 \quad \text{or} \quad (\gamma_5 \otimes (\gamma_5 \bar \gamma)^T) ,
\label{eq:d3_op}
\end{equation}
where $\bar \gamma \equiv \sum_{j=1}^3 \gamma_j$. Writing these possible
boundary contributions in terms of the one-component fields $\chi$ and
$\chib$ they then argue that the last two terms contain derivatives and are
of order $\cO(a)$ and therefore do not contribute in the continuum limit.
Staggered fermions, however, have additional symmetries, overlooked in
\cite{SF_KS}, namely shift invariance by one (fine) lattice spacing and a
charge conjugation symmetry (see the appendix A for a summary of the
symmetries of staggered fermions). Shift invariance (in spatial directions)
excludes the last two possibilities in (\ref{eq:d3_op}), making the more
indirect argument in \cite{SF_KS} unnecessary. $\tGt$, on the other hand,
is excluded by the charge conjugation symmetry. Thus the only possible
dimension three boundary counterterm is already present in the action. As
mentioned above it involves a boundary field and it can therefore be
absorbed into a renormalization of the boundary field, just as in the
continuum (or Wilson fermion) Schr\"odinger functional \cite{SF_Sint}.

\subsection{Boundary counterterms at order $a$}

Though the bulk part of the action, eq.~(\ref{eq:S_SF_psi}), appears to
have order $\cO(a)$ lattice effects --- the $a \Delta_\mu$ part of
$\cD_\mu$ --- this is not so. The apparent $\cO(a)$ effect just stems
from a ``bad'' choice in the construction of the four-component spinors
in eq.~(\ref{eq:psi_def}). It can be transformed away by using ``improved''
fields \cite{Luo}
\begin{equation}
\chi^I_\xi(y) = \chi_\xi(y) - a \sum_\nu \delta_{\xi_\nu,1} D_\nu
 \chi_\xi(y) , ~~ \chib^I_\xi(y) = \chib_\xi(y) - a \sum_\nu
 \delta_{\xi_\nu,1} \chib_\xi \lvec D_\nu(y) .
\end{equation}
For our Dirac spinors this transformation becomes
\begin{equation}
\psi^I(y) = \psi(y) - a \sum_\nu P_1^{(\nu)} \cdot D_\nu \psi(y) , ~~
 \psib^I(y) = \psib(y) - a \sum_\nu \psib \lvec D_\nu(y)
 \cdot P_1^{(\nu)} .
\label{eq:imp_psi_luo}
\end{equation}
This choice of improved fields is not unique. We prefer a somewhat more
symmetric form, which turns out to treat the fields near the boundaries in
the Schr\"odinger functional more equally,
\begin{eqnarray}
\psi^I(y) &=& \psi(y) - a \sum_\nu \half \left( P_1^{(\nu)} - P_0^{(\nu)}
 \right) \cdot D_\nu \psi(y) \nonumber \\
 \psib^I(y) &=& \psib(y) - a \sum_\nu \psib \lvec D_\nu(y)
 \cdot \half \left( P_1^{(\nu)} - P_0^{(\nu)} \right) .
\label{eq:imp_psi}
\end{eqnarray}
To apply the transformation, eq.~(\ref{eq:imp_psi}), near the boundaries it
is convenient to extend the fields $\psi$ and $\psib$ to the region $y_4 <
0$ and $y_4 > \Tp$ by setting them to zero there. Inserting the improved
fields into the action and doing a partial resummation one finds the
additional term in the bulk \cite{Luo}
\begin{displaymath}
- (2a)^4 a \sum_{y,\mu} \psib^I(y) D_\mu^2 \tGf_\mu \psi^I(y)
\end{displaymath}
which cancels the $\cO(a)$ term in eq.~(\ref{eq:S_SF_psi}), up to
higher orders in $a$. Luo \cite{Luo} shows in addition that no dimension
five operators (in the bulk) are allowed by the staggered symmetries and
hence that no bulk $\cO(a)$ artefacts will appear at the quantum level.

At the boundaries, however, $\cO(a)$ effects may occur. These come from
dimension four operators at the boundary. Operators allowed by the
staggered symmetries in the massless case under consideration are (see
appendix A for a review of those symmetries)
\begin{eqnarray}
\cO_1 &=& \sum_{k=1}^3 \psib \tG_k \cD_k \psi \nonumber \\
\cO_2 &=& \sum_{k=1}^3 \psib (\gamma_k \gamma_4 \gamma_5) \otimes
 (\gamma_5 \gamma_4)^T \cD_k \psi \\
\cO_3 &=& \psib(0) \tGt \frac{1}{a} \left[\psi(1) - \psi(0) \right] -
 \frac{1}{a} \left[\psib(1) - \psib(0) \right] \tGt \psi(0) \nonumber \\
\cO_4 &=& \psib(0) \tGft \frac{1}{a} \left[\psi(1) - \psi(0) \right] +
 \frac{1}{a} \left[\psib(1) - \psib(0) \right] \tGft \psi(0)
\label{eq:dim4_bd}
\end{eqnarray}
The term $\cO_1$ and the combination $\cO_3+\cO_4$ already occur in the
action, eq.~({\ref{eq:S_SF_psi}). $\cO_2$ and the combination
$\cO_3-\cO_4$, both flavor symmetry breaking --- they break the shift
symmetry in the time direction which is broken already, of course, by the
presence of the boundary in the Schr\"odinger functional --- appear at the
quantum level.

\subsection{Relation to the usual Schr\"odinger functional}

Starting from the transfer matrix for Wilson fermions, Sint arrived at
fermionic boundary conditions that are different from eq.~(\ref{eq:BC_hyp})
\cite{SF_Sint}.\footnote{Yet other boundary conditions were considered by
Symanzik \cite{SF_Sym}. However, those boundary conditions explicitly break
parity.} For the four-flavor Dirac spinors considered here they would read
\begin{equation}
P_+ \psi(\vy,0) = \rho(\vy) , \quad
\psib(\vy,0) P_- = \rhob(\vy) , \quad
P_- \psi(\vy,\Tp) = \rhop(\vy) , \quad
\psib(\vy,\Tp) P_+ = \rhobp(\vy) ,
\label{eq:BC_W}
\end{equation}
with $P_\pm = \half \left[ \tOne \pm \tGt \right]$. Hence the Schr\"odinger
functional with staggered fermions, defined thus far, does not seem to
agree with the ``usual'' definition of the Schr\"odinger functional in the
presence of fermions. However, Sint noted that the staggered action and
boundary conditions in the classical continuum limit derived in
\cite{SF_KS} could be brought into the form he derived for Wilson fermions
by a ``chiral rotation'' \cite{Sint_priv}
\begin{equation}
\psi^\prime = R \cdot \psi \quad \text{and} \quad
 \psib^\prime = \psib \cdot {\bar R}
\label{eq:sf_rot}
\end{equation}
where ${\bar R} = \tGt \cdot R^\dagger \cdot \tGt$ and
\begin{equation}
R = R^5_4 (\theta_5) = \exp\{i \theta_5 (i \tGft) \} = \cos \theta_5 \tOne
 - \sin \theta_5 \tGft = \bar R^5_4 (\theta_5) .
\label{eq:rot5}
\end{equation}
Defining
\begin{equation}
P_{0,1} (\theta_5) = R^5_4 (\theta_5) \cdot P_{0,1} \cdot
 (R^5_4 (\theta_5))^{-1} .
\end{equation}
one finds
\begin{equation}
P_{0,1} (\theta_5) = \half \left[ \tOne \pm \cos 2\theta_5 (\gamma_4 \gamma_5)
 \otimes (\gamma_5 \gamma_4)^T \mp \sin 2\theta_5 \tGt
 \right] .
\label{eq:proj_rot5}
\end{equation}
Hence one can smoothly go from the ``natural'' Schr\"odinger boundary
conditions for staggered fermions to the conventional ones for continuum
(and Wilson) fermions, at least in the massless case under discussion.
Since the rotation needed, eq.~(\ref{eq:rot5}), is chiral this will not be
true for massive fermions. We shall discuss massive fermions later on, in
section \ref{Schroed_m}.

The ``rotation'' (\ref{eq:sf_rot}) reflects an arbitrariness in the
assignment of Dirac and flavor indices in the construction of the
four-component spinors in eq.~(\ref{eq:psi_def}). Such a flavor-spinor
rotation could be inserted directly into (\ref{eq:psi_def}). An allowed
rotation should leave the kinetic term of the fermion action in the
continuum limit unchanged. Thus we require that
\begin{equation}
\bar R^{-1} \cdot \tG_\mu \cdot R^{-1} = \tG_\mu \quad \forall \mu .
\end{equation}
Therefore the generators of $R$ need to be
\begin{equation}
\GS \otimes \GFT = [{\bf 1}, \gamma_5] \otimes [{\bf 1}, \gamma_\mu,
\gamma_5, i(\gamma_5 \gamma_\mu), i(\gamma_\mu \gamma_\nu) ]^T .
\end{equation}
Note that all these generators are hermitian. In particular the generator
$i \tGft$, and hence the rotation $R_4^5$, (\ref{eq:rot5}), is allowed.

\section{The Schr\"odinger coupling in the massless case}
\label{coupling_m0}

The first real application of the Schr\"odinger functional formalism in
lattice gauge theory was the computation of a well--defined renormalized
coupling \cite{SF_1-SF_3}. The special boundary conditions allow
introduction of an external field to probe the system, and the finite size
of the system gives a definite scale. In this section we will compute
the one-loop contribution from the staggered fermions to this coupling and
show that it agrees with the result using (improved) Wilson fermions
\cite{SF_SW} or a continuum regularization. This serves as a test for the
correctness of the Schr\"odinger functional setup with staggered fermions.
As a by-product we will also obtain the contribution from the staggered
fermions to the pure gauge $\cO(a)$ boundary counterterm with
coefficient $c_t$ in the notation of \cite{SF_1,SF_3,SF_SW}.

The external field is introduced via the boundary gauge fields
taken as abelian with
\begin{eqnarray}
W_k(\vx) &=& {\rm diag} \left( {\rm e}^{i \phi_1/L} , {\rm e}^{i \phi_2/L} ,
 {\rm e}^{i \phi_3/L} \right) \nonumber \\
W^\prime_k(\vx) &=& {\rm diag} \left( {\rm e}^{i \phi^\prime_1/L} ,
 {\rm e}^{i \phi^\prime_2/L} , {\rm e}^{i \phi^\prime_3/L} \right) .
\end{eqnarray}
They lead to the classical gauge fields
\begin{equation}
U^{cl}_4(x) = 0 , \quad \left[ U^{cl}_k(\vx,x_4) \right]_{ij} = 
 \delta_{ij} {\rm e}^{i (x_4 \phi^\prime_j + (T-x_4) \phi_j)/(LT)} .
\end{equation}
As \cite{SF_3,SF_SW} we choose the boundary fields to depend on a
parameter, which we denote by $\omega$, through
\begin{eqnarray}
\phi_1 &=& -\frac{\pi}{3} + \omega , \quad \phi_2 = - \half \omega , \quad
 \phi_3 = \frac{\pi}{3} - \half \omega , \nonumber \\
\phi^\prime_1 &=& -\pi - \omega , \quad \phi^\prime_2 = \frac{\pi}{3}
 + \half \omega , \quad \phi^\prime_3 = \frac{2\pi}{3} + \half \omega .
\end{eqnarray}
The ``Schr\"odinger functional coupling constant'' is then defined as
\begin{equation}
\frac{k}{\bar g^2} = - \frac{\partial}{\partial \omega}
 \log {\mathcal Z} \Big|_{\omega=0}, \quad
 k = 12 \left( \frac{L}{a} \right)^2 \left[ \sin \left( \frac{2 \pi a^2}{3LT}
 \right) + \sin \left( \frac{\pi a^2}{3LT} \right) \right] ,
\label{eq:coupl_def}
\end{equation}
with $T=L$ such that $\bar g^2$ depends only on one scale, $\bar g^2 =
\bar g^2(L)$. The normalization $k$ has been chosen such that $\bar g$
equals the bare coupling at tree--level without any cutoff effects.

The one--loop contribution from the staggered fermions to the coupling
constant eq.~(\ref{eq:coupl_def}) comes from the derivative of the fermion
fluctuation determinant. The fermion boundary fields here are set to zero.
Then the fermion action can be written schematically as
\begin{equation}
S_{SF} = \sum_{\vx} \sum_{x_4=1}^{T-1} \sum_z \chib(x) M_{x,z} \chi(z) .
\end{equation}
As usual with staggered fermions, one easily sees that
\begin{equation}
(-1)^{|x|} M_{x,z} (-1)^{|z|} = M^\dagger_{z,x} .
\end{equation}
Thus $(-1)^{|x|} M_{x,z} \equiv {\mathcal M}_{x,z}$ is hermitian, and has
the same determinant as $M_{x,z}$. The one-loop contribution to the
running coupling is given by \cite{SF_SW}
\begin{equation}
\bar g^2 = g_0^2 + p_1 g_0^4 + \cO(g_0^6) , \quad p_1 = p_{1,0} + n_f
 p_{1,1} ,
\end{equation}
with the fermionic contribution
\begin{equation}
p_{1,1} = \frac{1}{k n_f} \frac{\partial}{\partial \omega} \log \det
 {\mathcal M} \Big|_{\omega=0} .
\label{eq:ferm_1_lp}
\end{equation}
Here $n_f=4$ for one flavor of staggered fermions, since they correspond to
four flavors of continuum fermions.

The eigenfunctions of ${\mathcal M}$ are of the form
\begin{equation}
\chi(x) = {\rm e}^{i (\vec p + \va \pi) \vx} f_{\va}(x_4) , \quad
 f_{\va}(x_4=0) = f_{\va}(x_4=T) = 0 ,
\end{equation}
where $\va$ is a 3-dimensional vector with $\alpha_k = 0,1$ and
\begin{equation}
p_k = \frac{2\pi n_k}{L} + \frac{\theta_k}{L} , \quad n_k=0,\ldots,\half
L_k -1 .
\end{equation}
As usual for staggered fermions the momentum components go only over half
the Brillouin zone interval. The remainder, $\alpha_k \pi$, becomes an
``internal'' index, loosely corresponding to spin/flavor.

Introducing $\eta^{(\mu)}$ as
\begin{equation}
\eta^{(\mu)}_\nu = \begin{cases}
 1 & \text{for $\nu < \mu$} \\
 0 & \text{for $\nu \geq \mu$} 
 \end{cases}
\end{equation}
such that the staggered phase factors become $\eta_\mu(x) = (-1)^{\eta^{(\mu)}
\cdot x}$ and realizing that $(-1)^{|x|} = (-1)^{\veta^{(4)} \vx + x_4}$ we
find that $f_{\va}(x_4)$ satisfies
\begin{eqnarray}
(-1)^{x_4} {\mathcal M}_{\va,\vb} f^{(i)}_{\vb}(x_4) &=&
 \half f^{(i)}_{\va}(x_4+1) - \sum_{k=1}^3 i \sin(r^{(i)}_k+\varphi^{(i)} x_4)
 (-1)^{\alpha_k} \bar \delta_{\va+\veta^{(4)}+\veta^{(\mu)},\vb}
 f^{(i)}_{\vb}(x_4) \nonumber \\
&-& \half f^{(i)}_{\va}(x_4-1) .
\label{eq:evol}
\end{eqnarray}
Here $\bar \delta$ means $\delta$ modulo 2, $r^{(i)}_k = p_k + \phi_i/L$
and $\varphi^{(i)} = (\phi^\prime_i - \phi_i)/(LT)$.

Essentially, due to the species doubling of staggered fermions in the time
direction, even though we are dealing with fermions, eq.~(\ref{eq:evol}) is
a (hermitian) second difference equation. We can therefore directly apply
the recursive method of appendix B of \cite{SF_1} to compute the
fluctuation determinant for fixed $\vec p$.

To be precise, the coupling is defined in \cite{SF_3,SF_SW} as
eq.~(\ref{eq:coupl_def}) with $T=L$ so that it depends on a single scale,
$L$. Unfortunately this is not possible for staggered fermions, since, as
we have seen, $L/a$ must be even but $T/a$ must be odd (we have restored
here the dimensions of $L$ and $T$).\footnote{The fermion fluctuation
determinant would be zero for $T/a$ even, yet another confirmation that
$T/a$ must be odd.} The closest we can get to $T=L$ is therefore $T=L \pm
a$. They all coincide in the continuum limit, but at finite $a$ taking $T=L
\pm a$ introduces additional $\cO(a)$ effects. Averaging the couplings
obtained with $T=L+a$ and $T=L-a$ cancels this additional $\cO(a)$ effect.

We have evaluated $p^{(\pm)}_{1,1}$ of eq.~(\ref{eq:ferm_1_lp}) for $L/a$
ranging from 4 to 64, in steps of 2. Here the superscript $\pm$ stands for
the choices $T=L \pm a$. One expects $p_{1,1}(L/a)$ (we omit the
superscript for the generic case) to be given by an asymptotic series of
the form \cite{SF_1,SF_SW}
\begin{equation}
p_{1,1}(L/a) = r_0 + s_0 \log(L/a) + (r_1 + s_1 \log(L/a) ) (a/L) +
 \cO(\log(L/a) (a/L)^2) .
\label{eq:p11_asym}
\end{equation}
The first few coefficients in eq.~(\ref{eq:p11_asym}) can be extracted by
first cancelling higher order terms in $a/L$ through numerical
differentiation and then checking for stability as $L/a$ is increased
\cite{LW86}. $s_0$, the coefficient of the logarithmically divergent term
in the continuum limit, should just be $2b_{0,1} = -1/(12\pi^2)$, the
fermionic contribution, per flavor, to the $\beta$-function \cite{SF_SW},
and thus absorbed by renormalization. We indeed found this result to about
6 digits accuracy for all cases considered.

$s_1$ was found to be compatible with zero, albeit only with an accuracy of
about 4 digits. To extract $r_0$ we assume the exact value for $s_0$; to
obtain $r_1$ we assume in addition that $s_1 = 0$. Then we found the values
listed in Table~\ref{tab:r0_m0}. $r_0$ obtained from the choices $T=L \pm
a$ always agreed to the accuracy given.

\begin{table}[htb]
\begin{center}
\begin{tabular}{|c|c|c|c|c|} \hline
 $\theta$ & $r_0$ & $r^{(+)}_1$ & $r^{(-)}_1$ & $r^{(av)}_1$ \\ \hline
 $0$     & $-0.004416(1)~~$ & $-0.02003(1)~$ & $0.03897(1)~$ &
    $0.00947(1)~$ \\
 $\pi/5$ & $-0.00579695(2)$ & $-0.023113(5)$ & $0.042061(5)$ &
    $0.009477(5)$ \\
 $1.0$   & $-0.0068642(1)~$ & $-0.02552(1)~$ & $0.04447(1)~$ &
    $0.00947(1)~$ \\
\hline
\end{tabular}
\end{center}
\caption{The first two `non-log' terms in the expansion
         eq.~(\protect\ref{eq:p11_asym}) of $p_{1,1}$. $r^{(\pm)}_1$ comes
         from the choices $T=L \pm a$ and $r^{(av)}_1$ is the average
         of the two.}
\label{tab:r0_m0}
\medskip \noindent
\end{table}

$r_0$ of eq.~(\ref{eq:p11_asym}) is the finite part of the fermionic
contribution to the one-loop relation between bare lattice and running
Schr\"odinger functional coupling. Using the known one-loop relation
between lattice and $\overline{\rm MS}$ coupling we can obtain a relation
between the Schr\"odinger functional coupling $\alpha$ and
$\alpha_{\overline{\rm MS}}$
\begin{equation}
\alpha_{\overline{\rm MS}}(q) = \alpha(q) + c_1 \alpha^2(q) + \cO(\alpha^3) ,
 \quad \alpha(q=1/L) = \bar g^2(L) / (4\pi) .
\end{equation}
This relation between two renormalized continuum couplings has to be
independent of the regularization used to obtain it. Thus, we should find
the same result for $c_{1,1}$ in $c_1 = c_{1,0} + n_f c_{1,1}$ as
\cite{SF_SW}. $c_{1,1}$ is given by
\begin{equation}
c_{1,1} = - 4 \pi [ P_4 + r_0 ]
\end{equation}
where $P_4 = 0.0026247371$ is the finite contribution from staggered
fermions to the one-loop relation between lattice and $\overline{\rm MS}$
coupling, computed in \cite{KS_ferm}.\footnote{The better accuracy quoted
here for $P_4$ is obtained from eq.~(6.12) of \cite{KS_ferm} using the more
accurate values for $P_1$ and $P_2$ from \cite{LW95}.} Using $r_0$ from
Table~\ref{tab:r0_m0} we obtain
\begin{equation}
c_{1,1} = \begin{cases}
 0.02251(1) & \text{for $\theta=0$} \\
 0.0398632(2) & \text{for $\theta=\pi/5$}
 \end{cases}
\end{equation}
in good agreement with the results of Sint and Sommer for Wilson fermions
\cite{SF_SW}. This confirms that our Schr\"odinger functional for massless
staggered fermions is a correct regularization of the continuum
Schr\"odinger functional with massless fermions.

\subsection{Lattice artefacts}

The vanishing of $s_1$ indicates the absence of bulk $\cO(a)$ artefacts
\cite{SF_SW}, as expected for staggered fermions \cite{Luo}. The
non-vanishing of $r_1$ reflects the presence of boundary $\cO(a)$ effects.
As mentioned before, the fact that we are not allowed to take $T=L$ for the
computation of the running coupling introduces additional $\cO(a)$ effects
that can not be cancelled by boundary counterterms. They can, however, be
cancelled by averaging over the choices $T=L \pm a$, as can be seen from
Table~\ref{tab:r0_m0}:\footnote{We expect this to be true also for the pure
gauge part that, in a simulation with dynamical staggered fermions, would
obviously have to come from an average of systems with $T=L \pm a$. We have
not checked this explicitly. But S.~Sint has checked that the statement
holds for the one-loop contribution from Wilson fermions. In addition we
have performed a pure gauge MC simulation for $\beta=5.9044$ with $L/a=4$
and $T = L \pm a$ and verified that the average of the inverse of the
non-perturbatively computed coupling constants agrees, within errors, with
the result for $T/a=L/a=4$ in \cite{SF_3}.} $r^{(av)}_1$ is independent of
the spatial boundary conditions parameterized by $\theta$. This remaining
$\cO(a)$ effect can be absorbed into the pure gauge boundary counterterm
$(c_t-1) \sum_{k=1}^3 {\rm tr}(F_{k4} F_{k4})$ \cite{SF_SW} by choosing
$c_t = 1 + (c_t^{(1,0)} + n_f c_t^{(1,1)}) g_0^2$ with the pure gauge part
$c_t^{(1,0)} = -0.08900(5)$ \cite{SF_3} and the fermionic contribution
\begin{equation}
c_t^{(1,1)} = \half r^{(av)}_1 = 0.00474(1) .
\label{eq:ct_ks}
\end{equation}

After cancelling the $\cO(a)$ boundary lattice artefact by the boundary
counterterm, higher order lattice artefacts, both from the bulk and the
boundary, remain. With our results we can study them for the fermionic
contribution to the step scaling function \cite{SF_3,SF_SW},
$\Sigma(s,u,a/L)$ which is the coupling $\bar g^2(sL)$ at scale $sL$ when
keeping the coupling at scale $L$ fixed at $\bar g^2(L)=u$. In the
continuum, the one-loop fermionic contribution to $\sigma(s,u) =
\Sigma(s,u,0)$ is $2b_{0,1} n_f \log(s) \ u^2$, while on the lattice it is
given by $n_f [p_{1,1}(2L/a) - p_{1,1}(L/a)] \ u^2$. We take $s=2$ and
compare the lattice result (per continuum flavor) --- we consider here the
average between the choices $T=L \pm a$ --- with its continuum limit:
\begin{equation}
\overline \delta_{1,1}(a/L) = \frac{p^{(av)}_{1,1}(2L/a) -
 p^{(av)}_{1,1}(L/a)}{2b_{0,1} \log2} .
\label{eq:delta_11}
\end{equation}
Deviation of $\overline \delta_{1,1}(a/L)$ from 1 at finite $a/L$ is a
lattice artefact. Note that we define the lattice artefact from the
fermions, $\overline \delta_{1,1}(a/L) -1$, with respect only to the
fermionic contribution in the continuum limit, in contrast to \cite{SF_SW}.
$\overline \delta_{1,1}(a/L)$ is shown in Figs.~\ref{fig:sb_0_0} and
\ref{fig:sb_pi5_0} for $\theta=0$ and $\pi/5$, respectively. Shown in both
cases is the result with and without the cancellation of the $\cO(a)$ part
of the lattice artefact by the pure gauge boundary counterterm. As can be
seen from the two cases, the higher order lattice artefacts can depend very
sensitively on the observable considered, here the step $\beta$-function
for different values of the spatial boundary conditions. After cancellation
of the $\cO(a)$ part by the boundary counterterm, one expect the remainder
to go asymptotically (up to logs) like $(a/L)^2$. Rough estimates of the
coefficient of the $(a/L)^2$ contribution from the figures are -7.5 and
-0.5 respectively. For $\theta=1$ the coefficient becomes about 4.5. These
rather large variations should serve as a caution to drawing conclusions
about the order of magnitude of cut-off effects from studying just one
observable.

\begin{figure}[htb]
  \vspace{3.6in}
  \includegraphics{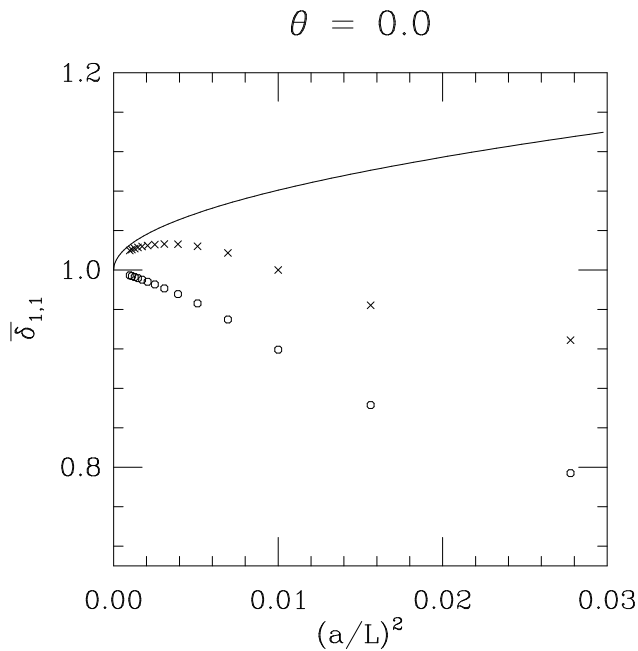}
  \caption{The ratio of the fermionic contribution to the step scaling
           function at one--loop from the lattice to its continuum limit,
           (\protect\ref{eq:delta_11}), for $\theta=0$. Crosses show the
           result obtained from the fermion fluctuation determinant without
           cancellation of the $\cO(a)$ part by the pure gauge boundary
           counterterm $c_t^{(1,1)}$, and octagons the result after the
           cancellation. The line shows the $\cO(a)$ part of the lattice
           artefact that is cancelled by the pure gauge boundary
           counterterm.}
  \label{fig:sb_0_0}
\end{figure}

\begin{figure}[htb]
  \vspace{3.6in}
  \includegraphics{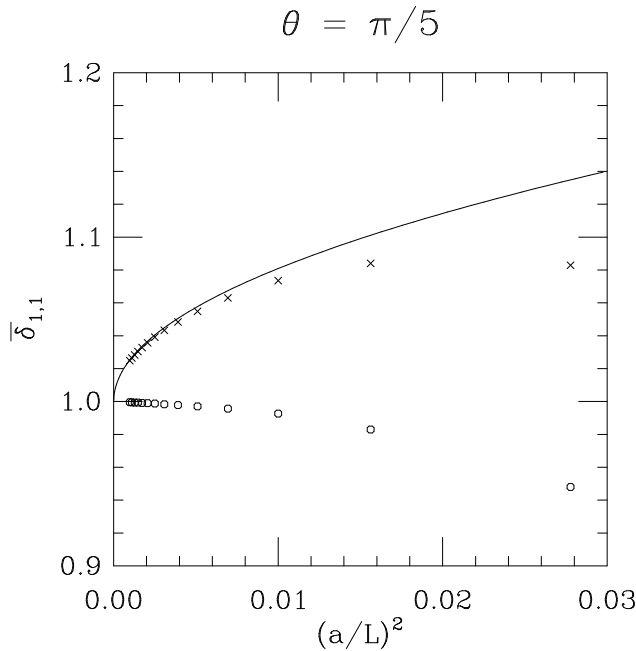}
  \caption{Same as Fig.~\protect\ref{fig:sb_0_0}, but for $\theta=\pi/5$.}
  \label{fig:sb_pi5_0}
\end{figure}

\section{The Schr\"odinger functional for massive staggered fermions}
\label{Schroed_m}

We have seen that with a chiral rotation, eq.~(\ref{eq:rot5}) with
$\theta_5 = \pi/4$, we can turn the ``natural'' staggered boundary
conditions, in terms of four-component Dirac spinors, into the usual
boundary conditions obtained when starting with Wilson fermions. Since this
is a chiral rotation, the mass term, omitted so far, will not remain
invariant. Thus, to obtain the usual Schr\"odinger functional boundary
conditions with the usual mass term we have two choices. Either we change
the boundary conditions for the staggered fermions, or we use an
unconventional mass term.

Here we consider the second alternative. Thus, instead of the usual mass
term $m \psib \psi$ in terms of the four-component spinors before the
chiral rotation we consider $m \psib \tGft \psi$, which becomes the usual
mass term {\it after} the chiral rotation. Using eq.~(\ref{eq:psi_def})
this mass term can be written in terms of the one-component spinors as:
\begin{equation}
S_{m_5} = m \sum_x \eta_4(x) \chib(x) \left\{ \half
 \left[1-(-1)^{x_4}\right] U^\dagger_4(x-\hat 4) \chi(x-\hat 4) - \half
 \left[1+(-1)^{x_4}\right] U_4(x) \chi(x+\hat 4) \right\} ,
\label{eq:m5_a}
\end{equation}
where we included the gauge fields to make this non-local (on the fine
lattice) mass term gauge invariant.

However, this new mass term, eq.~(\ref{eq:m5_a}), introduces an $\cO(a)$
term in the bulk. This can be seen by writing it in terms of the improved
four-component spinors, eq.~(\ref{eq:imp_psi}),
\begin{equation}
m (2a)^4 \sum_y \psib(y) \tGft \psi(y) = m (2a)^4 \sum_y \left[ \psib^I(y)
 \tGft \psi^I(y) + a \psib^I(y) \tGt D_4 \psi^I(y) \right] + \cO(a^2) .
\end{equation}
To the same order, we can replace $D_4$ in the second term by $\cD_4$.
Subtracting this term, we arrive at the $\cO(a)$ improved mass term
\begin{equation}
S_{m_5} = - m \sum_x \half (-1)^{x_4} \eta_4(x) \chib(x) \left(
 U_4(x) \chi(x+\hat 4) + U^\dagger_4(x-\hat 4) \chi(x-\hat 4) \right) .
\label{eq:m5_b}
\end{equation}

A few comments are in order here: the mass term (\ref{eq:m5_b}) obviously
singles out the time direction. But the time direction is singled out in
the Schr\"odinger functional formalism anyway. This mass term is invariant
under the ``chiral'' $U(1)_\epsilon$ symmetry, but it breaks the shift
symmetry in the time direction. Indeed, a shift by one lattice unit in the
time direction reverses the sign of the mass. Hence, for the usual setting
with (anti-) periodic boundary conditions, the model is invariant under a
change of sign of the mass. The free spectrum, for example, is a function
of $m^2$. However, such a shift in the time direction is not possible at
the boundary. The Schr\"odinger functional is thus sensitive to the sign of
the mass, and we therefore expect the Schr\"odinger coupling to depend
linearly on the mass for small masses. In the continuum limit, the shift
symmetries become part of the spin and vector and axial flavor symmetries.
Thus breaking of the shift symmetry implies breaking of the chiral symmetry
in the continuum, just as is expected of a mass term.

Since a mass term can break a symmetry, counterterms have to be re-examined
in its presence. The mass term (\ref{eq:m5_b}) only breaks the shift
symmetry in the time direction, but leaves the ``chiral'' $U(1)_\epsilon$
symmetry unbroken. In the bulk, the only allowed counterterm
multiplicatively renormalizes the mass. On the boundary,
the only $\cO(a)$ term allowed in addition to those in
eq.~(\ref{eq:dim4_bd}) is
\begin{equation}
\cO_5 = m \psib \tGft \psi ,
\label{eq:dim4_bd_m5}
\end{equation}
{\it i.e.,} the new term now already appearing in the action.

Having found a mass term that is designed to correspond to the usual mass
term in the Schr\"odinger functional formalism, we can check this at
one--loop by computing the contribution to the running coupling. The
computation goes as in the previous section. The eigenfunctions now satisfy
\begin{eqnarray}
(-1)^{x_4} {\mathcal M}_{\va,\vb} f^{(i)}_{\vb}(x_4) &=&
 \half \left[1+(-1)^{x_4}m \right] f^{(i)}_{\va}(x_4+1) \nonumber \\
&-& \sum_{k=1}^3 i \sin(r^{(i)}_k+\varphi^{(i)} x_4) (-1)^{\alpha_k}
 \bar \delta_{\va+\veta^{(4)}+\veta^{(\mu)},\vb} f^{(i)}_{\vb}(x_4) \\
&-& \half \left[1-(-1)^{x_4}m \right] f^{(i)}_{\va}(x_4-1) . \nonumber
\label{eq:evol_m5}
\end{eqnarray}

Since the scale for the running coupling is given by the spatial system
size $L$ we want to keep $z=mL$ fixed when varying $L/a$. Again, we compute
the contribution $p_{1,1}(z,L/a)$ for $L/a =4$ up to 64 and extract the
leading terms in the asymptotic expansion eq.~(\ref{eq:p11_asym}). A few
examples are listed in Table~\ref{tab:r0_m5}. Again we find $s_0 =
-1/(12\pi^2)$ to our accuracy, and $s_1$ compatible with zero.\footnote{We
actually discovered the presence of a bulk $\cO(a)$ contribution for the
unimproved mass term, eq.~(\ref{eq:m5_a}), by finding a non-vanishing
$s_1$.}

\begin{table}[htb]
\begin{center}
\begin{tabular}{|c|c|c|c|c|c|} \hline
 $\theta$ & $z$ & $r_0$ & $r^{(+)}_1$ & $r^{(-)}_1$ & $r^{(av)}_1$ \\ \hline
 $0$     & $1.0$ & $0.0024408(1)~$ & $-0.00426(1)$ & $0.02319(1)$ &
    $0.00947(1)$ \\
 $\pi/5$ & $1.0$ & $0.00179605(3)$ & $-0.00617(1)$ & $0.02511(1)$ &
    $0.00947(1)$ \\
 $0$     & $2.0$ & $0.0064310(1)~$ & $~0.00223(1)$ & $0.01671(1)$ &
    $0.00947(1)$ \\
 $\pi/5$ & $2.0$ & $0.00632882(5)$ & $~0.00203(1)$ & $0.01691(1)$ &
    $0.00947(1)$ \\
\hline
\end{tabular}
\end{center}
\caption{$r_0$ and $r_1$'s as in Table~\protect\ref{tab:r0_m0} but in the
         massive case with $z=mL$ held fixed.}
\label{tab:r0_m5}
\medskip \noindent
\end{table}

The difference $c_{1,1}(z) - c_{1,1}(0) = -4\pi[r_0(z) - r_0(0)]$ should be
regularization independent. Comparing with the Wilson fermion results of
\cite{SF_SW} we indeed find agreement within the accuracy given in Tables
\ref{tab:r0_m0} and \ref{tab:r0_m5}.

The value of $r^{(av)}_1$ is independent of $z$ and hence cancelled by the
pure gauge boundary counterterm with coefficient (\ref{eq:ct_ks}). No mass
dependent $\cO(a)$ bulk term of the form $m {\rm tr}(F_{\mu \nu}
F_{\mu \nu})$ is needed, in contrast to the case of Wilson fermions
\cite{SF_SW}.

If we do not insist on reproducing the conventional ({\it i.e.} Wilson
fermion inspired) massive Schr\"odinger functional in the continuum limit,
we can use the ``natural'' staggered boundary conditions and the usual
(degenerate) staggered mass term\footnote{We do not consider mass terms
that break some of the staggered symmetries (other than the chiral
symmetry) which might be used to lift the degeneracy among the four
continuum flavors.} without chiral rotation
\begin{equation}
S_m = m \sum_x \chib(x) \chi(x) = (2a)^4 m \sum_y \psib(y) \psi(y) .
\label{eq:Sm}
\end{equation}
This leads to a well--defined Schr\"odinger functional in the continuum,
albeit with boundary conditions that mix the four continuum flavors. In
contrast to the ``conventional'' case now both action and boundary
conditions are invariant under $m \to -m$ and thus the Schr\"odinger
functional running coupling defined by these conventions is an even
function of $z=mL$. Inclusion of the mass term (\ref{eq:Sm}) into
eq.~(\ref{eq:evol}) is straightforward, giving
\begin{eqnarray}
(-1)^{x_4} {\mathcal M}_{\va,\vb} f^{(i)}_{\vb}(x_4) &=&
 \half f^{(i)}_{\va}(x_4+1) - \sum_{k=1}^3 i \sin(r^{(i)}_k+\varphi^{(i)} x_4)
 (-1)^{\alpha_k} \bar \delta_{\va+\veta^{(4)}+\veta^{(\mu)},\vb}
 f^{(i)}_{\vb}(x_4) \nonumber \\
&+& m \bar \delta_{\va+\veta^{(4)},\vb} f^{(i)}_{\vb}(x_4)
 - \half f^{(i)}_{\va}(x_4-1) .
\label{eq:evol_m}
\end{eqnarray}
Proceeding as before, we now obtain the sample results listed in
Table~\ref{tab:r0_m}. We note in particular that $r^{(av)}_1$ is unchanged
from before and hence cancelled by the pure gauge boundary counterterm with
coefficient (\ref{eq:ct_ks}).

\begin{table}[htb]
\begin{center}
\begin{tabular}{|c|c|c|c|c|c|} \hline
 $\theta$ & $z^2$ & $r_0$ & $r^{(+)}_1$ & $r^{(-)}_1$ & $r^{(av)}_1$ \\ \hline
 $0$     & $1.0$ & $-0.0004189(1)~$ & $-0.00853(1)~$ & $0.02747(1)~$ &
    $0.00947(1)~$ \\
 $\pi/5$ & $1.0$ & $-0.00195976(3)$ & $-0.013405(3)$ & $0.032350(3)$ &
    $0.009473(3)$ \\
\hline
\end{tabular}
\end{center}
\caption{$r_0$ and $r_1$'s as in Table~\protect\ref{tab:r0_m0} but in the
         massive case with the standard staggered mass term,
         (\protect\ref{eq:Sm}).}
\label{tab:r0_m}
\medskip \noindent
\end{table}

The mass term (\ref{eq:Sm}) breaks the ``chiral'' $U(1)_\epsilon$
symmetry. As a consequence the $\cO(a)$ boundary counterterms
\begin{eqnarray}
\cO_6 &=& m \psib \psi \nonumber \\
\cO_7 &=& m \psib (\gamma_4 \gamma_5) \otimes (\gamma_5 \gamma_4)^T \psi .
\end{eqnarray}
are allowed in addition to those in eq.~(\ref{eq:dim4_bd}) and the term
in (\ref{eq:dim4_bd_m5}).

\section{Conclusions}
\label{conclusions}

We have discussed the Schr\"odinger functional with staggered fermions,
extending and completing previous work by Miyazaki and Kikukawa. In
particular, we have shown that for massless fermions the Schr\"odinger
functional constructed agrees, in the continuum limit, with the
conventional Schr\"odinger functional for fermions as described by Sint, by
examining the Schr\"odinger functional for staggered fermions in terms of
four flavors of Dirac spinors. We found one boundary counterterm that
contributes in the continuum limit, just as for Wilson fermions and for
continuum fermions with dimensional regularization. It can be absorbed by a
renormalization of the boundary fields.

We have computed the one-loop contribution from massless staggered fermions
to the running Schr\"odinger functional coupling and found agreement, in
the continuum limit, with the results of Sint and Sommer for Wilson
fermions. We computed the one--loop contribution from staggered fermions to
the pure gauge $\cO(a)$ boundary counterterm, which cancels the $\cO(a)$
lattice artefact in the running coupling. We discussed the higher order
lattice artefacts in the step scaling function at one loop. They can be as
large as 20\%  on an $L/a=6$ lattice, and they can depend rather
sensitively on the observable considered, in our case the coupling defined
with different fermionic boundary conditions.

A chiral, flavor changing rotation was needed to bring the boundary
condition of staggered fermions, expressed in terms of four flavors of
Dirac fermions, to the conventional form. The mass term is not invariant
under such a rotation. We therefore considered an unconventional mass term
for the staggered fermions, constructed to give the conventional
Schr\"odinger functional for massive fermions in the continuum limit. The
construction was verified by computing the one-loop contribution to the
Schr\"odinger coupling constant for fixed $z=mL$ and reproducing results by
Sint and Sommer.

One unpleasantness in defining the Schr\"odinger functional running
coupling constant with dynamical staggered fermions is the fact that we
cannot take $T=L$, since $L/a$ has to be even while $T/a$ has to be odd.
Therefore the coupling does not strictly depend on a single scale. Taking
$T = L \pm a$ gives a single scale in the continuum limit ($a \to 0$), but
introduces $\cO(a)$ effects at finite lattice spacing. As discussed, these
$\cO(a)$ effects, which can not be cancelled by boundary counterterms, can
be cancelled by averaging over the coupling obtained with $T = L \pm a$.
However, since simulations with staggered fermions tend to be less costly
than simulations with Wilson-type fermions --- staggered fermions have four
times fewer degrees of freedom, and no fine--tuning is required to make
them massless --- having to do two simulations for each $L/a$ might not be
such a big price to pay.

Having established the Schr\"odinger functional with dynamical staggered
quarks opens the possibility to use it for the non-perturbative computation
of improvement coefficients and current renormalization constants for gauge
field ensembles with dynamical staggered fermions, just as it is done for
quenched simulations by the ALPHA collaboration \cite{SF_cSW,SF_Z}. We
envision here the computation of the clover coefficient and current
renormalization constants for (improved) Wilson valence quarks. These can
then be used to compute phenomenologically interesting quantities, such as
{\it e.g} $f_B$ and $f_D$, with Wilson valence fermions on gauge
configurations generated with dynamical staggered fermions
\cite{HEMCGC_ME}. Comparison with quenched results then allows an
estimation of quenching errors \cite{MILC_fB} in those quenched
computations.

\section*{Acknowledgements}

This research was supported by DOE contracts DE-FG05-85ER250000
and DE-FG05-96ER40979. We thank S.~Sint for numerous discussions and
suggestions, and for providing the program to extract the leading
coefficients in the asymptotic expansion eq.~(\ref{eq:p11_asym}).
We also thank Y.~Kikukawa for helpful comments on an early draft of this
manuscript and T.~Kennedy for a critical reading of it.

\section*{Appendix A}
\renewcommand{\theequation}{A.\arabic{equation}}

In this appendix we briefly review the symmetries of staggered fermions
\cite{Golt_Smit,Jolicoeur}. Unless otherwise noted, the one-component
spinors transform under a symmetry transformation $T$ as
\begin{eqnarray}
\chi(x) &\to& \chi^\prime(x) = \eta_T(T^{-1} x) \chi(T^{-1} x) \nonumber \\
\chib(x) &\to& \chib^\prime(x) = \eta_T(T^{-1} x) \chib(T^{-1} x) .
\label{eq:transf_chi}
\end{eqnarray}
The gauge fields transform as
\begin{equation}
U(x,z) \to U(T^{-1} x, T^{-1} z)
\end{equation}
with
\begin{equation}
U(x,z) = \begin{cases}
 U_\mu(x) & \text{for $z=x+\mu$} \\
 U^\dagger_\mu(z) & \text{for $x=z+\mu$.}
 \end{cases}
\end{equation}

\vspace{0.5cm}
\noindent (i) Rotations by $\pi/2$:

Here we consider rotations around the center of a hypercube ($\rho <
\sigma$):
\begin{eqnarray}
R_H^{\rho,\sigma}: \quad x_\rho &\to& x_\sigma \nonumber \\
 x_\sigma &\to& - x_\rho + 1 \\
 x_\mu &\to& x_\mu \quad \text{for $\mu \ne \rho, \sigma$} . \nonumber
\end{eqnarray}
Then we have
\begin{equation}
\eta_{R_H}(x) (-1)^{(x_\rho+x_\sigma)(x_{\rho+1}+\ldots+x_{\sigma-1})
 + x_\rho(x_\sigma+1) + x_{\sigma+1}+\ldots+x_4} .
\end{equation}
In terms of the hypercube block coordinates:
\begin{equation}
y_\rho \to y_\sigma , \quad y_\sigma \to -y_\rho , \quad y_\mu \to y_\mu
 \quad \text{for $\mu \ne \rho, \sigma$} ,
\end{equation}
and the fields $\psi$ and $\psib$ transform as
\begin{eqnarray}
\psi(y) &\to& \half (1 + \gamma_\rho \gamma_\sigma) \otimes
 (\gamma_\sigma^T - \gamma_\rho^T) \cdot \psi(R_H^{-1}y) \nonumber \\
\psib(y) &\to& \psib(R_H^{-1}y) \cdot \half (1 - \gamma_\rho \gamma_\sigma)
 \otimes (\gamma_\sigma^T - \gamma_\rho^T) .
\end{eqnarray}

\vspace{0.5cm}
\noindent (ii) Reflections on hyperplanes:

Here we consider reflections on hyperplanes through the center of a
hypercube:
\begin{eqnarray}
I_H^{\rho}: \quad x_\rho &\to& - x_\rho + 1 \nonumber \\
 x_\mu &\to& x_\mu \quad \text{for $\mu \ne \rho$} .
\end{eqnarray}
Then we have
\begin{equation}
\eta_{I_H}(x) (-1)^{x_\rho +\ldots +x_4} .
\end{equation}
In terms of the hypercube block coordinates:
\begin{equation}
y_\rho \to -y_\rho , \quad y_\mu \to y_\mu \quad \text{for $\mu \ne \rho$} ,
\end{equation}
and the fields $\psi$ and $\psib$ transform as
\begin{equation}
\psi(y) \to (\gamma_\rho \gamma_5) \otimes \gamma_5^T \cdot
 \psi(I_H^{-1}y) , \quad
\psib(y) \to \psib(I_H^{-1}y) \cdot (\gamma_5 \gamma_\rho) \otimes
 \gamma_5^T .
\end{equation}

Combining 3 reflections, one orthogonal to each spatial direction gives the
parity transformation
\begin{equation}
P: \quad \vy \to -\vy , \quad y_4 \to y_4 ,
\end{equation}
with
\begin{equation}
\psi(y) \to \gamma_4 \otimes \gamma_5^T \cdot \psi(Py) , \quad
\psib(y) \to \psib(Py) \cdot \gamma_4 \otimes \gamma_5^T .
\end{equation}

\vspace{0.5cm}
\noindent (iii) Shift invariance:

\begin{eqnarray}
T^{\rho}: \quad x_\rho &\to& x_\rho + 1 \nonumber \\
 x_\mu &\to& x_\mu \quad \text{for $\mu \ne \rho$} .
\end{eqnarray}
Then we have
\begin{equation}
\eta_{T}(x) (-1)^{x_{\rho+1} +\ldots +x_4} ,
\end{equation}
and the fields $\psi$ and $\psib$ transform as
\begin{eqnarray}
\! \! \! \! \! \! \! \! \! \! \psi(y) & \! \! \! \to \! \! \! &
 \half \delta_{y,y^\prime} (1 \otimes \gamma_\rho^T -
 \gamma_\rho \gamma_5 \otimes \gamma_5^T ) \cdot \psi(y^\prime) +
 \half \delta_{y+\rho,y^\prime} (1 \otimes \gamma_\rho^T +
 \gamma_\rho \gamma_5 \otimes \gamma_5^T ) \cdot \psi(y^\prime) \nonumber \\
\! \! \! \! \! \! \! \! \! \! \psib(y) & \! \! \! \to \! \! \! &
 \half \delta_{y,y^\prime} \psib(y^\prime) \cdot
 (1 \otimes \gamma_\rho^T + \gamma_\rho \gamma_5 \otimes \gamma_5^T ) +
 \half \delta_{y,y^\prime-\rho} \psib(y^\prime) \cdot (1 \otimes
 \gamma_\rho^T - \gamma_\rho \gamma_5 \otimes \gamma_5^T ) .
\end{eqnarray}

Symmetries (i) through (iii) are the space-time symmetries of staggered
fermions. In addition we have the invariances:

\vspace{0.5cm}
\noindent (iv) U(1)-invariance:

\begin{equation}
\chi(x) \to {\rm e}^{i \alpha} \chi(x) , \quad
\chib(x) \to {\rm e}^{-i \alpha} \chib(x) ,
\end{equation}
which just becomes
\begin{equation}
\psi(y) \to {\rm e}^{i \alpha} \psi(y) , \quad
\psib(y) \to {\rm e}^{-i \alpha} \psib(y) .
\end{equation}

and

\vspace{0.5cm}
\noindent (v) U(1)$_\epsilon$-invariance:

\begin{equation}
\chi(x) \to {\rm e}^{i \beta \epsilon(x)} \chi(x) , \quad
\chib(x) \to {\rm e}^{i \beta \epsilon(x)} \chib(x) ,
\end{equation}
where $\epsilon(x) = (-1)^{|x|}$. This becomes the ``chiral''
transformation
\begin{equation}
\psi(y) \to {\rm e}^{i \beta \gamma_5 \otimes \gamma_5^T} \cdot \psi(y) , \quad
\psib(y) \to \psib(y) \cdot {\rm e}^{i \beta \gamma_5 \otimes \gamma_5^T} .
\end{equation}
This chiral symmetry protects the zero-mass limit for staggered fermions,
since the usual mass term $m \sum_x \chib(x) \chi(x) = m 2^4 \sum_y
\psib(y) \psi(y)$ is not invariant.

Finally, staggered fermions have the discrete

\vspace{0.5cm}
\noindent (vi) Interchange symmetry:

\begin{equation}
\chi(x) \to \epsilon(x) \chib^T(x) , \quad
\chib(x) \to - \chi^T(x) \epsilon(x) ,
\end{equation}
where $T$ stands for transpose (as a color vector). The gauge fields
transform as
\begin{equation}
U_\mu(x) \to U^\ast_\mu(x) .
\end{equation}
In terms of the four-component spinors this becomes charge conjugation
symmetry
\begin{equation}
\psi(y) \to {\mathcal C} \psib^T(y) , \quad
\psib(y) \to - \psi^T(y) {\mathcal C} .
\end{equation}
Here ${\mathcal C} = C \otimes (C^{-1})^T$ where $C$ is the usual Euclidean
charge conjugation symmetry matrix satisfying
\begin{eqnarray}
C \gamma_\mu C^{-1} &=& -\gamma_\mu^T \nonumber \\
C \gamma_5 C^{-1} &=& \gamma_5^T \nonumber \\
C \gamma_5 \gamma_\mu C^{-1} &=& (\gamma_5 \gamma_\mu)^T \\
-C = C^T &=& C^{-1} = C^\dagger .
\end{eqnarray}

\section*{Appendix B}
\renewcommand{\theequation}{B.\arabic{equation}}

In this appendix we sketch the construction of a ``two--time--slice''
transfer matrix for staggered fermions, following quite closely the
construction in \cite{KS_ferm} for the ``reduced'' staggered fermions. We
consider staggered fermions with the standard (flavor degenerate) mass
term. The partition function is
\begin{equation}
Z = \int [DU] [d\chib d\chi] \exp\{-S_G - S_F\} .
\label{eq:Z_KS}
\end{equation}
with $S_G$ the usual Wilson gauge action and
\begin{equation}
S_{F} = \sum_{x} \sum_\mu \half \eta_\mu(x)
 \chib(x) \left[ U_\mu(x) \chi(x+\mu) - U^\dagger_\mu(x-\mu)
 \chi(x-\mu) \right] + \sum_x m \chib(x) \chi(x) .
\label{eq:S_Fm}
\end{equation}
We now change integration variables for each pair of time slices $x_4 =
2\tau$ and $2\tau+1$
\begin{eqnarray}
\alpha_\tau^\dagger(\vx) &=& P_e(\vx) \half \eta_4(\vx) \chib(\vx,2\tau)
 + P_o(\vx) \chi(\vx,2\tau) \nonumber \\
\alpha_\tau(\vx) &=& P_e(\vx) \chi(\vx,2\tau+1)
 + P_o(\vx) \half \eta_4(\vx) \chib(\vx,2\tau+1) \nonumber \\
\beta_\tau^\dagger(\vx) &=& P_e(\vx) \chi(\vx,2\tau)
 + P_o(\vx) \half \eta_4(\vx) \chib(\vx,2\tau) \\
\beta_\tau(\vx) &=& P_e(\vx) \half \eta_4(\vx) \chib(\vx,2\tau+1)
 + P_o(\vx) \chi(\vx,2\tau+1) \nonumber
\end{eqnarray}
where $P_{e,o}(\vx) = \half \left( 1 \pm (-1)^{|\vx|} \right)$ are the
projectors onto even and odd spatial sites $\vx$. We have used here that
staggered phase factors $\eta_\mu$ are independent of the time coordinate;
in particular, $\eta_4(\vx,x_4) = \eta_4(\vx) = (-1)^{|\vx|}$.

Introducing
\begin{displaymath}
\teta_k(\vx) = \eta_4(\vx) \eta_k(\vx) , \qquad
\teta_k(\vx+\hk) = -\teta_k(\vx) ,
\end{displaymath}
we can write the staggered action in temporal gauge, $U_4(x)=1$, as
\begin{eqnarray}
S_F &=& \sum_\tau \sum_\vx \left\{ \left[\alpha_\tau^\dagger(\vx)
 \alpha_\tau(\vx) - \alpha_{\tau+1}^\dagger(\vx) \alpha_\tau(\vx) 
 + \beta_\tau^\dagger(\vx) \beta_\tau(\vx)
 - \beta_{\tau+1}^\dagger(\vx) \beta_\tau(\vx) \right] \right. \nonumber \\
&+& P_e(\vx) \sum_k \left[ \alpha_\tau^\dagger(\vx) \teta_k(\vx)
 U_k(\vx,2\tau) \alpha_\tau^\dagger(\vx+\hk)
 + \beta_\tau^\dagger(\vx+\hk) \teta_k(\vx) U_k^\dagger(\vx,2\tau)
 \beta_\tau^\dagger(\vx) \right. \nonumber \\
& & \left. \qquad + \alpha_\tau(\vx+\hk) \teta_k(\vx)
 U_k^\dagger(\vx,2\tau+1) \alpha_\tau(\vx)
 + \beta_\tau(\vx) \teta_k(\vx) U_k(\vx,2\tau+1)
 \beta_\tau(\vx+\hk) \right] \nonumber \\
&+& P_o(\vx) \sum_k \left[ \beta_\tau^\dagger(\vx) \teta_k(\vx)
 U_k(\vx,2\tau) \beta_\tau^\dagger(\vx+\hk)
 + \alpha_\tau^\dagger(\vx+\hk) \teta_k(\vx) U_k^\dagger(\vx,2\tau)
 \alpha_\tau^\dagger(\vx) \right. \nonumber \\
& & \left. \qquad + \beta_\tau(\vx+\hk) \teta_k(\vx)
 U_k^\dagger(\vx,2\tau+1) \beta_\tau(\vx)
 + \alpha_\tau(\vx) \teta_k(\vx) U_k(\vx,2\tau+1)
 \alpha_\tau(\vx+\hk) \right] \nonumber \\
&+& \left. 2m \left[ \alpha_\tau^\dagger(\vx) \beta_\tau^\dagger(\vx)
 + \beta_\tau(\vx) \alpha_\tau(\vx) \right] \right\} .
\label{eq:S_Fm_new}
\end{eqnarray}

Next we introduce operators $\hat \alpha(\vx)$, $\hat \alpha^\dagger(\vx)$,
$\hat \beta(\vx)$ and $\hat \beta^\dagger(\vx)$ with the anticommutation
relations
\begin{equation}
\{ \hat \alpha(\vx), \hat \alpha^\dagger(\vx^\prime) \}
 = \{ \hat \beta(\vx), \hat \beta^\dagger(\vx^\prime) \}
 = \delta_{\vx,\vx^\prime} 
\label{eq:anti_com}
\end{equation}
and all other anticommutators vanishing. On the Fock space, spanned by
these operators, we consider the coherent states
\begin{equation}
\Bigl| \alpha_\tau, \beta_\tau \Big\rangle \equiv \exp \left\{ \sum_\vx
 \left[ \hat \alpha^\dagger(\vx) \alpha_\tau(\vx)
 + \hat \beta^\dagger(\vx) \beta_\tau(\vx) \right] \right\}
 \bigl| 0 \big\rangle
\end{equation}
and
\begin{equation}
\Big\langle \alpha_\tau^\dagger, \beta_\tau^\dagger \Bigr| \equiv
 \big\langle 0 \bigr| \exp \left\{ \sum_\vx
 \left[ \alpha_\tau^\dagger(\vx) \hat \alpha(\vx)
 + \beta_\tau^\dagger(\vx) \hat \beta(\vx) \right] \right\}
\end{equation}
where $| 0 \rangle$ is the Fock vacuum. It can be shown that these coherent
states satisfy the completeness relation
\begin{equation}
\mathbf{1} = \int \left[d \alpha_\tau^\dagger d \alpha_\tau \right]
 \left[d \beta_\tau^\dagger d \beta_\tau \right]
 \exp \left\{ - \sum_\vx \left[ \alpha_\tau^\dagger(\vx) \alpha_\tau(\vx)
 + \beta_\tau^\dagger(\vx) \beta_\tau(\vx) \right] \right\}
 \Bigl| \alpha_\tau, \beta_\tau \Big\rangle 
 \Big\langle \alpha_\tau^\dagger, \beta_\tau^\dagger \Bigr| ,
\label{eq:comp_rel}
\end{equation}
and that for a normal ordered operator $\hat A = :A( \hat \alpha^\dagger,
\hat \beta^\dagger; \hat \alpha, \hat \beta):$, where normal ordering means
that all daggered operators appear to the left of all non-daggered
operators,
\begin{equation}
\Big\langle \alpha_{\tau+1}^\dagger, \beta_{\tau+1}^\dagger \Bigr| \hat A
 \Bigl| \alpha_\tau, \beta_\tau \Big\rangle = A( \alpha_{\tau+1}^\dagger,
 \beta_{\tau+1}^\dagger; \alpha_\tau, \beta_\tau)
 \exp \left\{ \sum_\vx \left[ \alpha_{\tau+1}^\dagger(\vx) \alpha_\tau(\vx)
 + \beta_{\tau+1}^\dagger(\vx) \beta_\tau(\vx) \right] \right\} .
\label{eq:mat_elem}
\end{equation}

We can see that the terms in the exponentials in (\ref{eq:comp_rel}) and
(\ref{eq:mat_elem}) reproduce the terms from the first line of the action
$S_F$, eq.~(\ref{eq:S_Fm_new}). The other terms can be reproduced from the
transfer matrix (in temporal gauge)
\begin{equation}
\hat \mathbf{T} = \hat \mathbf{T}_G^{1/2} \hat \mathbf{T}_F^\dagger
 \hat \mathbf{T}_G \hat \mathbf{T}_F \hat \mathbf{T}_G^{1/2}
\label{eq:T_mat}
\end{equation}
with $\hat \mathbf{T}_G$ the pure gauge transfer matrix (in temporal gauge)
\cite{L_C} --- we need a total of $\hat \mathbf{T}_G^2$ since we move two
time slices forward. For $\hat \mathbf{T}_F$ we find
\begin{eqnarray}
\hat \mathbf{T}_F &=& \exp \left\{ - \sum_\vx \left( P_e(\vx) \left[
 \hat \alpha(\vx+\hk) \teta_k(\vx) \hat U_k^\dagger(\vx) \hat \alpha(\vx)
 + \hat \beta(\vx) \teta_k(\vx) \hat U_k(\vx) \hat \beta(\vx+\hk)
 \right] \right. \right. \nonumber \\
& & \qquad \qquad + P_o(\vx) \left[ \hat \beta(\vx+\hk) \teta_k(\vx)
 \hat U_k^\dagger(\vx) \hat \beta(\vx) + \hat \alpha(\vx) \teta_k(\vx)
 \hat U_k(\vx) \hat \alpha(\vx+\hk) \right] \nonumber \\
& & \left. \left. \qquad \qquad + m \hat \beta(\vx) \hat \alpha(\vx)
 \right) \right\} .
\label{eq:T_mat_F}
\end{eqnarray}
With some algebra one can show that $\hat \mathbf{T}_F^\dagger
\hat \mathbf{T}_F$ is equal to the square of the ``one--time--slice''
transfer matrix of \cite{KS_ferm} when gauge fields and the mass term are
included in the latter.

The ``two--time--slice'' transfer matrix $\hat \mathbf{T}$,
(\ref{eq:T_mat}), is self--adjoint and positive. Therefore, restoring the
lattice spacing in the time direction, $a_t$, we can define a Hamiltonian
by
\begin{equation}
\hat \mathbf{T} = \mathrm{e}^{- 2 a_t \hat \mathbf{H} } .
\label{eq:H_def}
\end{equation}
$\hat \mathbf{H}$ is a complicated function of the fermion and gauge field
operators. It simplifies in the naive time continuum limit, $a_t \to 0$,
where we can neglect contributions of $\cO(a_t^2)$ and higher. The
fermionic part becomes in this limit
\begin{eqnarray}
\hat H_F &=& \sum_\vx \left\{ \half P_e(\vx) \left[ \hat \alpha^\dagger(\vx)
 \teta_k(\vx) \hat U_k(\vx) \hat \alpha^\dagger(\vx+\hk)
 + \hat \alpha(\vx+\hk) \teta_k(\vx) \hat U_k^\dagger(\vx) \hat \alpha(\vx)
 \right. \right. \nonumber \\
& & \left. \qquad \qquad + \hat \beta^\dagger(\vx+\hk) \teta_k(\vx)
 \hat U_k^\dagger(\vx) \hat \beta^\dagger(\vx)
 + \hat \beta(\vx) \teta_k(\vx) \hat U_k(\vx) \hat \beta(\vx+\hk)
 \right] \nonumber \\
&+& \qquad \half P_o(\vx) \left[ \hat \beta^\dagger(\vx) \teta_k(\vx)
 \hat U_k(\vx) \hat \beta^\dagger(\vx+\hk)
 + \hat \beta(\vx+\hk) \teta_k(\vx) \hat U_k^\dagger(\vx)
 \hat \beta(\vx) \right. \\
& & \left. \qquad \qquad + \hat \alpha^\dagger(\vx+\hk) \teta_k(\vx)
 \hat U_k^\dagger(\vx) \hat \alpha^\dagger(\vx)
 + \hat \alpha(\vx) \teta_k(\vx) \hat U_k(\vx) \hat \alpha(\vx+\hk)
 \right] \nonumber \\
&+& \left. \qquad m \left[ \hat \alpha^\dagger(\vx) \hat \beta^\dagger(\vx) 
 + \hat \beta(\vx) \hat \alpha(\vx) \right] \right\} . \nonumber
\label{eq:H_F_a}
\end{eqnarray}
We can bring this Hamiltonian to a more conventional form with the
canonical transformation
\begin{eqnarray}
\hat \chi(\vx) &=& \frac{1}{\sqrt{2}} P_e(\vx) \left(
 \hat \beta^\dagger(\vx) - \hat \alpha(\vx) \right)
 + \frac{1}{\sqrt{2}} P_0(\vx) \left( \hat \beta(\vx)
 - \hat \alpha^\dagger(\vx) \right) \nonumber \\
\hat \psi(\vx) &=& \frac{1}{\sqrt{2}} P_e(\vx) \left(
 \hat \beta^\dagger(\vx) + \hat \alpha(\vx) \right)
 + \frac{1}{\sqrt{2}} P_0(\vx) \left( \hat \beta(\vx)
 + \hat \alpha^\dagger(\vx) \right) .
\end{eqnarray}
The operators $\hat \chi$ and $\hat \psi$ and their adjoint's $\hat
\chi^\dagger$ and $\hat \psi^\dagger$ satisfy anticommutation relations like
eq.~(\ref{eq:anti_com}). Expressed in terms of these operators the
Hamiltonian (\ref{eq:H_F_a}) reads
\begin{eqnarray}
\hat H_F &=& \half \sum_{\vx,k} \left[ \hat \psi^\dagger(\vx)
 \teta_k(\vx) \hat U_k(\vx) \hat \psi^\dagger(\vx+\hk)
 + \hat \psi(\vx+\hk) \teta_k(\vx) \hat U_k^\dagger(\vx) \hat \psi(\vx)
 \right. \nonumber \\
& & \left. \qquad  + \hat \chi(\vx) \teta_k(\vx) \hat U_k(\vx)
 \hat \chi(\vx+\hk) + \hat \chi^\dagger(\vx+\hk) \teta_k(\vx)
 \hat U_k^\dagger(\vx) \hat \chi^\dagger(\vx) \right] \\
&+& m \sum_\vx (-1)^{|\vx|} \left[ \hat \psi^\dagger(\vx) \hat \psi(\vx) 
 - \hat \chi^\dagger(\vx) \hat \chi(\vx) \right] . \nonumber
\label{eq:H_F_b}
\end{eqnarray}
We recognize (\ref{eq:H_F_b}) as the Susskind Hamiltonian \cite{Suss77} for
two independent staggered fermion fields $\psi$ and $\chi$.

\end{document}